\documentclass[aps,preprint,showpacs,floatfix]{revtex4-1}
\usepackage{graphicx,bm,amsmath,dcolumn}
\begin{document}
\newcommand{\be}{\begin{equation}}
\newcommand{\ee}{\end{equation}}
\newcommand{\bea}{\begin{eqnarray}}
\newcommand{\eea}{\end{eqnarray}}
\newcommand{\half}{\frac{1}{2}}
\newcommand{\ith}{^{(i)}}
\newcommand{\im}{^{(i-1)}}
\newcommand{\np}
{\hbox{\lower0.5ex\hbox{$-$}\llap{\raise0.5ex\hbox{$+$}}}}
\newcommand{\pn}
{\hbox{\lower0.5ex\hbox{$+$}\llap{\raise0.5ex\hbox{$-$}}}}
\newcommand{\nn}
{\hbox{\lower0.5ex\hbox{$-$}\llap{\raise0.5ex\hbox{$-$}}}}
\newcommand{\pp}
{\hbox{\lower0.5ex\hbox{$+$}\llap{\raise0.5ex\hbox{$+$}}}}
\newcommand{\gae}
{\,\hbox{\lower0.5ex\hbox{$\sim$}\llap{\raise0.5ex\hbox{$>$}}}\,}
\newcommand{\lae}
{\,\hbox{\lower0.5ex\hbox{$\sim$}\llap{\raise0.5ex\hbox{$<$}}}\,}
\newcommand{\mat}[1]{{\bf #1}}
\title{Special transitions in an O($n$) loop model with an Ising-like
constraint}
\author{Zhe Fu$^{1}$, Wenan Guo$^{2}$ and Henk W. J. Bl\"ote$^{3}$ } 
\affiliation{$^{1}$College of Physics and Electronic Engineering, 
Xinxiang University, Xinxiang 453003, P. R. China}
\affiliation{$^{2}$Physics Department, Beijing Normal University,
Beijing 100875, P. R. China}
\email[Corresponding author: ]{waguo@bnu.edu.cn}
\affiliation{$^{3}$ Instituut Lorentz, Leiden University,
P.O. Box 9506, 2300 RA Leiden, The Netherlands}
\email{henk@lorentz.leidenuniv.nl}
\date{\today} 
\begin{abstract}
We investigate the O($n$) nonintersecting loop model on the square
lattice  under the constraint that the loops consist of ninety-degree
bends only. 
The model is governed by the loop weight $n$, a weight $x$ for each
vertex of the lattice visited once by a loop, and a weight $z$ for each 
vertex visited twice by a loop. We explore the $(x,z)$ phase diagram
for some values of $n$. For $0<n<1$, the diagram has the same topology as
the generic O($n$) phase diagram with $n<2$, with a first-order line
when $z$ starts to dominate, and an O($n$)-like transition when $x$
starts to dominate. Both lines meet in an exactly solved  higher
critical point. For
$n>1$, the O($n$)-like transition line appears to be absent. 
Thus, for $z=0$, the $(n,x)$ phase diagram displays a line of phase
transitions for $n\le 1$. The line ends at $n=1$ in an infinite-order
transition. We determine the conformal anomaly and the critical
exponents along this line. These results agree accurately with a recent
proposal for the universal classification of this type of model,
at least in most of the range $-1 \leq n \leq 1$.
We also determine the exponent describing crossover to the
generic O($n$) universality class, by introducing topological
defects associated with the introduction of `straight' vertices
violating the ninety-degree-bend rule.
These results are obtained by means of transfer-matrix calculations and
finite-size scaling.
\end{abstract}
\maketitle 

\section{Introduction}
\label{intro}
\begin{figure}
\includegraphics[scale=0.40]{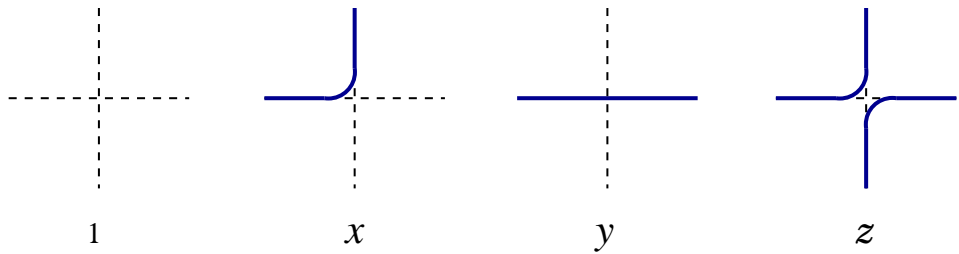}
\centering
\caption{(Color online)
The four kinds of vertices of the O($n$) loop model on the square lattice,
together with their weights. Rotated versions have the same weights.
The present work is mostly restricted to the subspace $(x,0,z)$ of
the $(x,y,z)$ model, with a special focus on the $(x,0,0)$ subspace.
}
\label{vertices}
\end{figure}

The present work investigates the nonintersecting loop model described by
the partition sum
\begin{equation}
Z_{{\rm loop}} =\sum_{{\rm all} \; {\mathcal G}}
x^{N_x} y^{N_y} z^{N_z} n^{N_l}, 
\label{Zloop}
\end{equation}
where ${\mathcal G}$ is a graph consisting of any number of $N_l$ closed,
nonintersecting loops. 
Each lattice edge may be covered by at most one loop segment, and  
there can be 0, 2, or 4 incoming loop segments at a vertex. In the
latter case, they can be connected in two different ways without having 
intersections. The allowed four kinds of vertices configurations are
shown in Fig.~\ref{vertices}, together with their weights denoted $x$,
$y$ and $z$.  The numbers of vertices with these weights are denoted
$N_x$, $N_y$, $N_z$ respectively.

A number of such loop models in two dimensions is exactly
solvable \cite{N,Baxter,BB,BNW,3WBN,KNB,3WPSN,VF,S,PS,GNB}.
In Ref.~\onlinecite{BNW}, five branches of critical points were found,
one of which describes the densely packed loop phase, and a second
branch describes its critical transition to a dilute loop gas. The
latter branch describes the generic O($n$) critical behavior,
and corresponds precisely with a result found earlier for the
honeycomb O($n$) model \cite{N}. A fifth branch found in 
Ref.~\onlinecite{BN}, called branch 0, is of particular interest
for the present work as a special case in the $y=0$ subspace.

\begin{figure}
\includegraphics[scale=0.50]{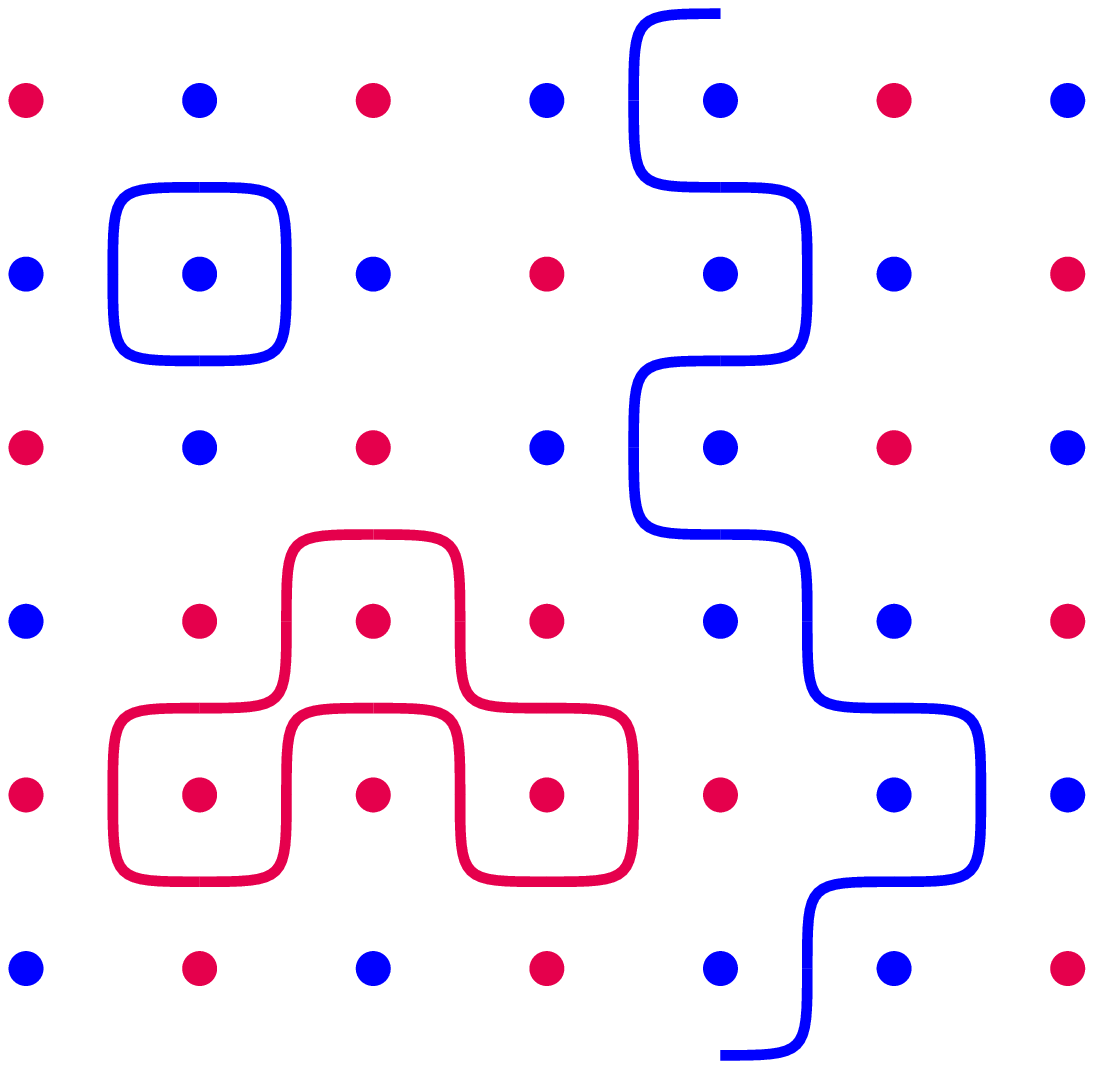} \hspace{15mm}
\includegraphics[scale=0.50]{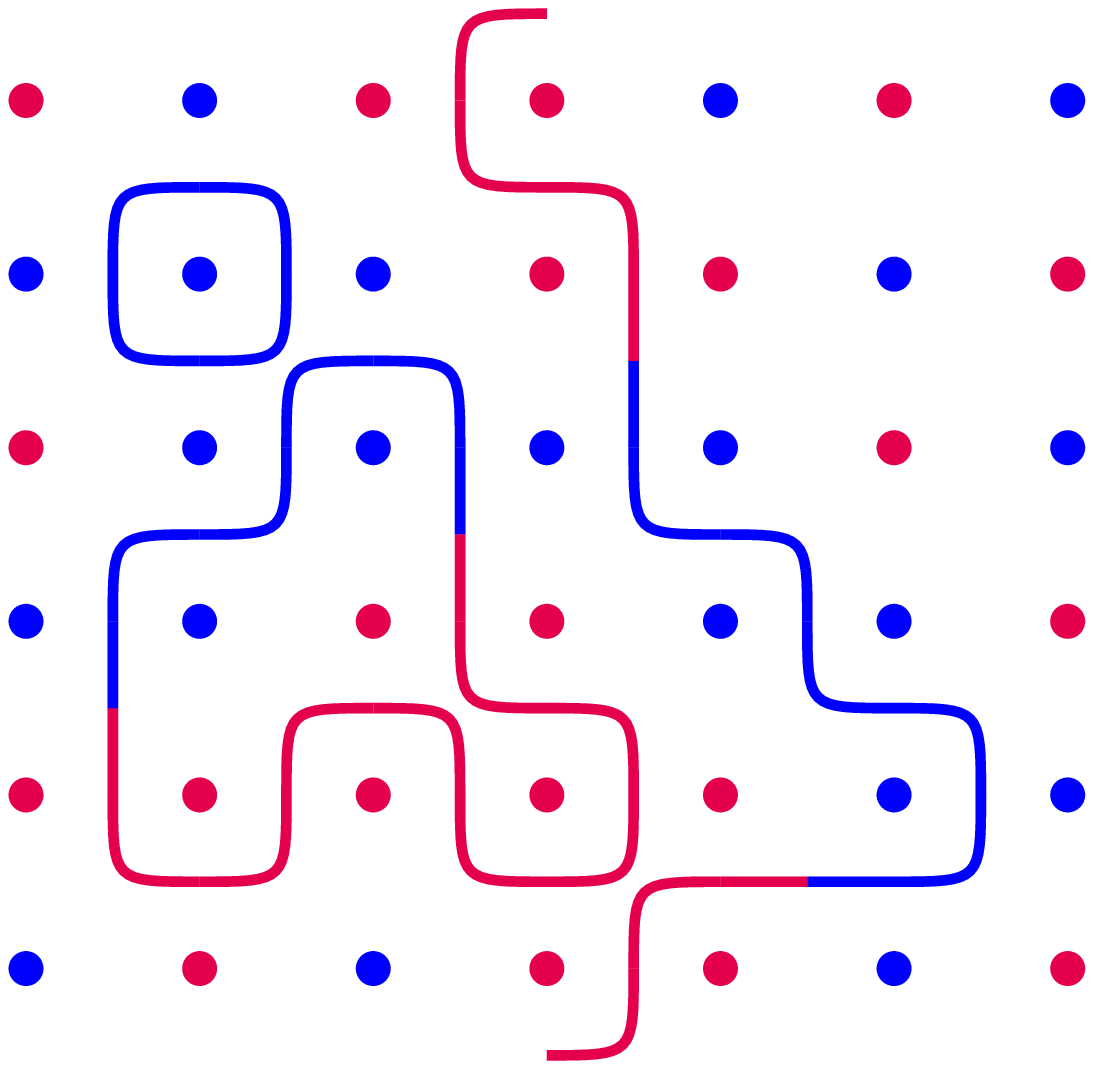}
\centering
\caption{(Color online)
Illustration of the Ising degree of freedom of O($n$) loops on the square
lattice. Dual neighbor spins have opposite signs, unless separated by a loop.
If $y$-type vertices are absent (left-hand side), each loop has a
single Ising color. The presence of $y$-type vertices (right-hand side)
leads to a change of sign of the Ising variable.
}
\label{ising}
\end{figure}

It is known that this generic behavior 
of the square-lattice O($n$) loop model can be modified
by Ising-like degrees of freedom of the loop configurations \cite{BN}.
These degrees of freedom are exposed by placing dual Ising spins 
$\pm 1$ on the faces of the lattice, with the rule that nearest
neighbors are of the same sign if and only if separated by a loop.
Figure \ref{ising} illustrates that each $y$-type vertex corresponds with
a change of sign of this Ising variable.

Suppression of the $y$-type vertex freezes the Ising degree of freedom
of each separate loop. Thus, in the case $n=0$, where we have at most
one loop, we may expect the generic O(0) behavior. For other values of
$n$, the Ising degrees of freedom of adjacent loops can be different,
which may influence the way they interact, and thereby modify their
universal behavior. This was indeed found in earlier work \cite{FGB},
using numerical investigations of the $y=0$ case. Vernier {\em et al.}
\cite{VJS} proposed the universal classification of this type of models
as that of the generic O(2$n$) behavior. Furthermore, the phase diagram
is modified for $y=0$. This will be demonstrated by the phase diagrams in
the $x,z$ plane for $n=1/2$ and $n=3/2$, presented in Sec.~\ref{phdiag}.
In Sec.~\ref{numres} we present numerical results for the
conformal anomaly and the magnetic and temperature scaling dimensions.
These agree well with the O(2$n$) classification, in particular for $n$ 
not too large.  Section \ref{numres} also includes a
determination of the topological dimension $X_y$ governing the crossover
from the $y=0$ model to the generic O($n$) model, and a proposal for its
universal classification.

A summary of the transfer-matrix technique is given in Sec.~\ref{transfmat},
including some remarks on the coding of the $y=0$ connectivities,
which allows us to obtain results up to finite size $L=20$. The paper
concludes with a short discussion in Sec.~\ref{disc}.

\section{The transfer-matrix analysis}
\label{transfmat}
We consider a square O($n$) model wrapped on an cylinder with one
set of edges in the length direction. The partition function of 
such a system with a circumference $L$ and a sufficiently large length
$M$, expressed in lattice units, satisfies 
\begin{equation}
Z(M,L)\simeq \Lambda_0(L)^M \, ,
\label{Z}
\end{equation}
where $\Lambda_0(L)$ is the largest eigenvalue of the transfer matrix.
A derivation of this formula for the present case of nonlocal interactions
is given {\em e.g.}, in Refs.~\onlinecite{BN82,WGB}. The transfer
matrix indices are numbers that refer to ``connectivities'', namely
the way that the dangling loop segments are pairwise connected when
one cuts the cylinder perpendicular to its axis.
The transfer matrix technique used here is in principle the same as
described in Ref.~\onlinecite{BN82}, except the coding of the 
``connectivities'' defined there. In principle one can use the coding
of the O($n$) loop connectivities described in Ref.~\onlinecite{BN}.
However, the special case $y=0$ opens the possibility of a more
efficient coding.

The evaluation of the largest eigenvalues of the transfer matrix is
done numerically. The size of the transfer matrix increases exponentially
with $L$, so that only a limited range of finite sizes can be handled.
Our calculations are limited to transfer matrix sizes up to about
$10^8 \times 10^8$ corresponding with $L \leq 20$.

Apart from the leading eigenvalue $\Lambda_0(L)$, we still determine
the second largest one $\Lambda_1(L)$. We also consider the case of a 
single loop segment running in the length direction of the cylinder,
which actually leads to a different set of connectivities, and another
sector of the transfer matrix. Its largest eigenvalue is denoted
$\Lambda_2(L)$.

\subsection{Use of the eigenspectrum}
From Eq.~(\ref{Z}) we obtain the free energy density as
\begin{equation}
f(L)=L^{-1} \ln\Lambda_{0}(L) \, .
\label{flt}
\end{equation}
The numerical results for $f(L)$ can be used to estimate the
conformal anomaly $c$ \cite{BCN,Affl} from
\begin{equation}
f(L) \simeq f+ \frac{\pi c}{ 6 L^{2} } \, .
\label{fcp}
\end{equation}
To avoid complications associated with alternation effects between
even and odd system sizes, the present numerical work is mainly focused 
on even system sizes.

The gap between $\Lambda_0(L)$ and the subleading eigenvalues is
used to determine the thermal and magnetic correlation lengths.
These quantities are expressed as scaled gaps $X_t$ and $X_h$
\begin{equation}
X_t(L)= \frac{L}{2\pi} \ln \frac{\Lambda_0 }{| \Lambda_1 |}\, ,
~~~~~~
X_h(L)= \frac{L}{2\pi} \ln \frac{\Lambda_0 }{| \Lambda_2 |}\, .
\label{Xth}
\end{equation}
The finite-size results for the scaled gaps yield estimates of the
scaling dimensions \cite{JCxi}:
\begin{equation}
X_t=\lim_{L \to \infty} X_t(L)\, , ~~~~~~ X_h=\lim_{L \to \infty} X_h(L)\, .
\label{xisce}
\end{equation}
These calculations are restricted to translationally invariant
(zero-momentum) eigenstates of the transfer matrix.

\subsection{Coulomb gas results}
For the generic critical O($n$) model in two dimensions, the conformal
anomaly $c$ is known \cite{BCN,BB} to be equal to
\begin{equation}
c=1-\frac{6(1-g)^2}{g}, ~~~~2\cos(\pi g)=-n, ~~~~1 \leq g \leq 2 \, .
\label{caCG}
\end{equation}
This range of $g$ corresponds with the critical O($n$) phase transition,
but the same formula with $0 \leq g \leq 1$ applies to dense phase.
The scaling dimensions $X_t$ and $X_h$ of the generic O($n$) model are
also known, see Ref.~\onlinecite{CG} and references therein:
\begin{equation}
X_t=\frac{4}{g} -2\, , ~~~~~~~ X_h=\frac{g}{8}-\frac{1}{2g}(1-g)^2
\label{XthCG}
\end{equation}
The exponent of the leading correction to scaling in the critical 
O($n$) model was also obtained with the Coulomb gas method \cite{CG}:
\begin{equation}
X_u= 2g-\frac{1}{2g}(1-g)^2 \, ,   
\label{Xu}
\end{equation}

\subsection{Method of analysis}
\label{metan}
From Eq.~(\ref{fcp}) one may estimate the conformal anomaly from
subsequent finite-size results $f(L)$ and $f(L+2)$ as
\begin{equation}
c(L)\equiv[f(L)-f(L+2)]\, \frac{3L^2(L+2)^2}{2 \pi (L+1)} \, .
\end{equation}
Taking into account corrections to scaling with exponent $y_u$,
these estimates are expected to behave as 
\begin{equation}
c(L)\simeq c+ b L^{y_u} ~~ {\rm with} ~~ y_u=2-X_u \, .
\end{equation}
The estimation of $c$ from the $f(L)$ is done on the basis of these
two formulas and three-point fits, as described {\em e.g.}, in
Refs.~\onlinecite{BN82,FSS}. The scaling dimensions are estimated
similarly from the scaled gaps defined above.

\subsection{Coding for the $y=0$ case}
\label{y0coding}

The transfer-matrix algorithm applied in Ref.~\onlinecite{FGB} used
the full set of well-nested O($n$) connectivities, {\em i.e.}, the
set corresponding with nonintersecting loops. However, for $y=0$, 
there is only a restricted set of O($n$) connectivities. If the $k$th
and the $m$th edges at the end of the cylinder are occupied by dangling
segments of the same loop, then $k-m$ is restricted to be odd in
the absence of straight $y$-type vertices (we consider only the case
of even $L$). This restriction considerably
reduces the number of allowed connectivities, with more than a factor
ten for the largest system size used. We wrote a new coding-decoding
algorithm for this case, thus obtaining a large reduction of the size
of the transfer matrix. This enabled us to handle somewhat larger
systems for $y=0$ than those in past numerical studies for $y \ne 0$.
~~\\

\subsection{The special case $n=1$}
For $n=1$, the transfer matrix simplifies because the weights depend
only on the number of loop segments, and not on the number of loops.
We represent the loops by dual Ising spins $\pm 1$ such that nearest
neighbors are of different signs if and only if separated by a loop.
After assigning local 4-spin Ising weights $W(\pp\pp)=W(\nn\nn)=1$,
$W(\pn\pp)=x$, $W(\pn\np)=2z$, etc., one reproduces the O(1) vertex
weights. Then one can easily apply a simple Ising transfer matrix,
and handle system sizes up to $L=28$.

\section{Numerical results}
\label{numres}
The results presented in Secs.~\ref{phdiag} and \ref{crpts} include
phase transitions that were located on the basis of the asymptotic
finite-size-scaling equation 
\begin{equation}
X_h(x,L) \simeq X_h(x,L+2)\, .
\label{Xhllp}
\end{equation}
The vertex weight $x$ was solved numerically, with the parameters
$z$ and $n$ kept constant. These solutions were denoted $x_c(L)$.
Best estimates of $x_c$ were obtained after extrapolation with a
procedure outlined in Ref.~\onlinecite{BN82}. Depending on the slope
of a phase transition line in the $x,z$ plane, one may solve for $z$
instead while keeping $x$ constant. At $x=0$, the exact locations $z_c$
of the transitions follow by equating the free energy of the vacuum state
to that of the completely packed state, {\em i.e.}, $z_c=\exp[-f(n)]$
where $f(n)$ is the free energy density of the completely packed model
with $z=1$. The function $f(n)$ was already found by Lieb \cite{Lieb}
for an equivalent 6-vertex model; for further details, see {\em e.g.}
Ref.~\onlinecite{BWG}.  Another exactly known
critical point is the branch-0 point \cite{BN} at $x=z=1/2$.

In Sec.~\ref{phdiag} we present the $x,z$ phase diagram for  a few
values of $n$. The subsections  thereafter concern the estimation of the
critical points and universal quantities as a function of $n$ along
the critical line for $z=0$.

\subsection{Phase diagrams for $n=0.5$ and $1.5$}
\label{phdiag}
\begin{figure}
\includegraphics[scale=1.10]{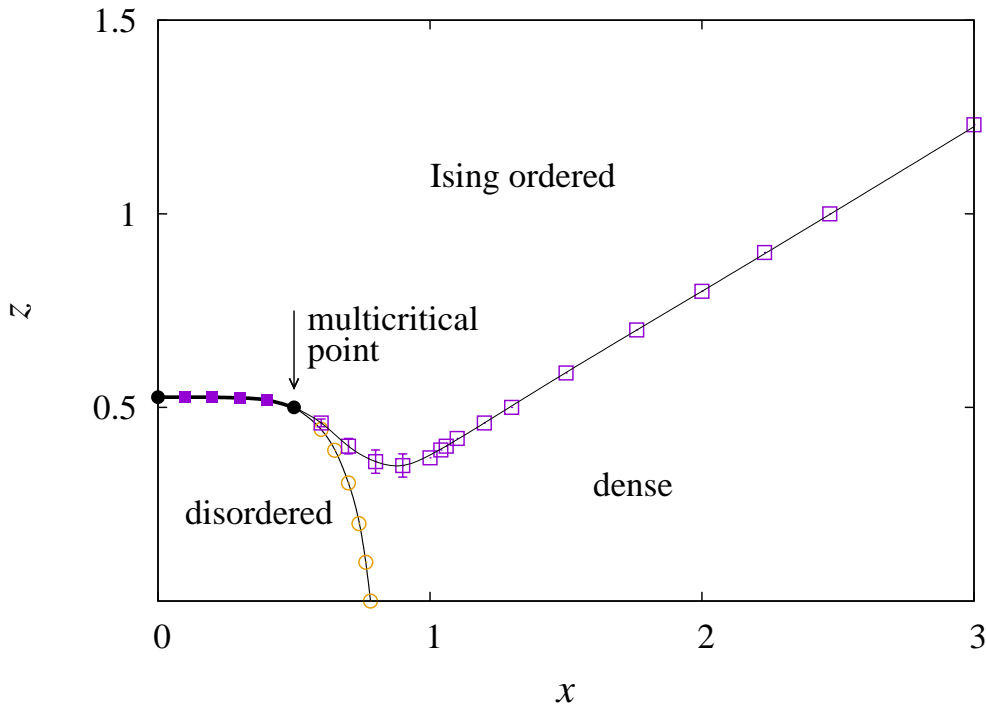}
\centering
\caption{(Color online)
Phase diagram in the $x,z$ plane for $n=1/2$.
The critical O($n$) line, which separates the disordered phase from the
dense  O($n$) phase, is seen to merge with an Ising critical line in an
exactly solved  multicritical point $x=z=1/2$. For $x$ smaller than the
multicritical value, the line of Ising transitions continues as a
first-order line, ending at the exactly known point $x=0$, 
$z=0.52652729\cdots$. The other data points were numerically obtained.
The curves serve only as a guide to the eye.}
\label{n0p5}
\end{figure}

\begin{figure}
\includegraphics[scale=1.10]{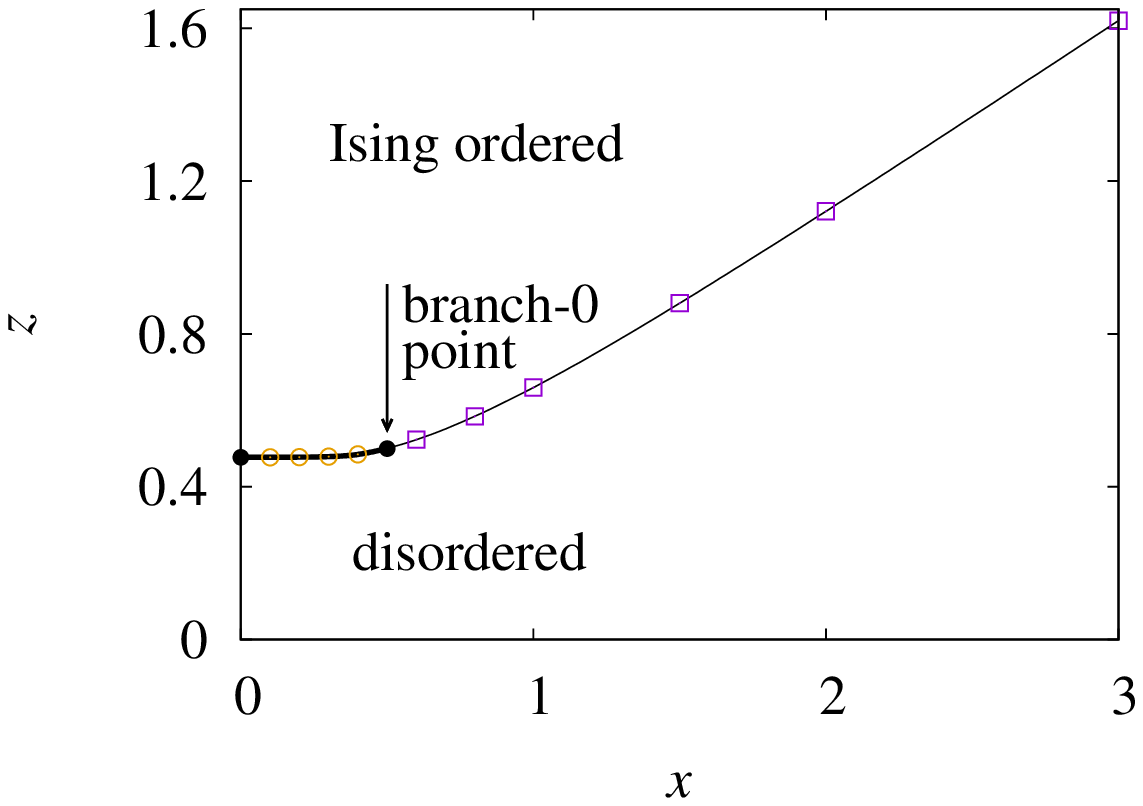}
\centering
\caption{(Color online)
Location of the Ising ordering transition 
in the $(x,z)$ plane of the square-lattice O(1.5) loop model.     
Exactly known  points are shown as black circles, the other data points
are numerically determined.
}
\label{n1p5xz}
\end{figure}
In both figures (Figs.~\ref{n0p5} and ~\ref{n1p5xz}) one observes a
first-order line coming in horizontally
on the vertical axis, separating the disordered phase from an Ising 
ordered phase. For $n=0.5$ one also observes a line of critical points
where the largest loops diverge.  This critical line meets the first-order
line in a multicritical point called ``branch 0'' in Ref.~\onlinecite{BN}.
A separate O($n$) critical line appears to be absent in the  O(1.5) model.
In both figures, the line of Ising-like transitions continues beyond the
branch 0-point. The Ising-like ordered phase at larger $z$ is dominated 
by $z$-type vertices, and the majority of the dual Ising spins are
antiferromagnetically ordered.

Although the Ising disordered phase at larger $x$ is labeled ``dense''
in Fig.~\ref{n0p5}, it is different from the dense phase such as
described in Refs.~\onlinecite{N,BN} because the individual loops are
already Ising ordered.

\begin{figure}
\includegraphics[scale=1.10]{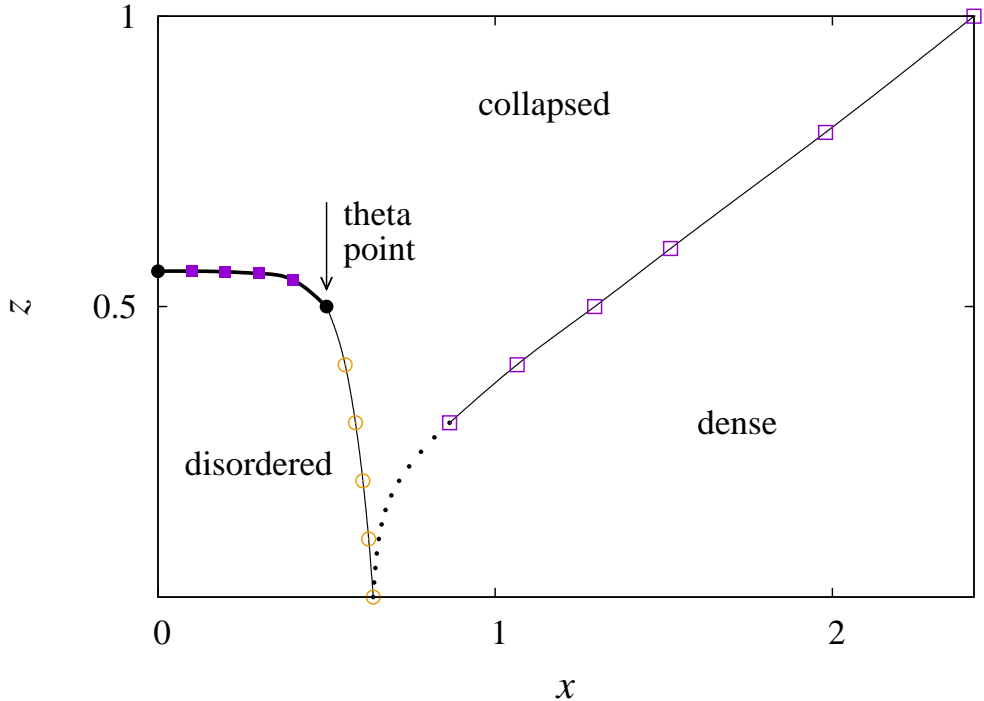}
\centering
\caption{(Color online)
Phase diagram in the $x,z$ plane for $n=0$.
The O($0$) transition line on the left-hand side consists of a first-order
part (thick line) and a continuous transition line going down to the
$x$ axis. There is also a transition line on the right-hand side, which
could be clearly located down to $z=0.3$. While the part with $z<0.3$,
shown as small dots, is less accurately determined, the behavior of
the magnetic gaps suggests that it continues to $z=0$.
}
\label{n0p0}
\end{figure}

The phase diagram for the special case $n=0$  was already
investigated \cite{BBN} some time ago. Since there is at most one loop
which is already Ising ordered, a nonzero density of that loop leads
directly to a nonzero staggered magnetization of the dual spins.
The question may arise if there is still an Ising-like transition
for $x>0.5$ when $z$ increases. But the scaled magnetic gaps display
clear intersections, indicating that there is still a phase transition
line on the right-hand side of the $n=0$ diagram. This line
was not observed in Ref.~\onlinecite{BBN}, which focused on the 
the O(0) transition line and the branch-0 point $x=z=1/2$ which was
identified as a $\theta$ point describing a collapsing polymer.
The numerical analysis becomes difficult in the neighborhood of
$x=0.7$ for small $z$. Our interpretation, shown in Fig.~\ref{n0p0},
is that the transition line goes to $z=0$ when it approaches the O(0)
line, while its critical amplitudes vanish.

\subsection{Critical points}
\label{crpts}
\begin{figure}
\includegraphics[scale=1.10]{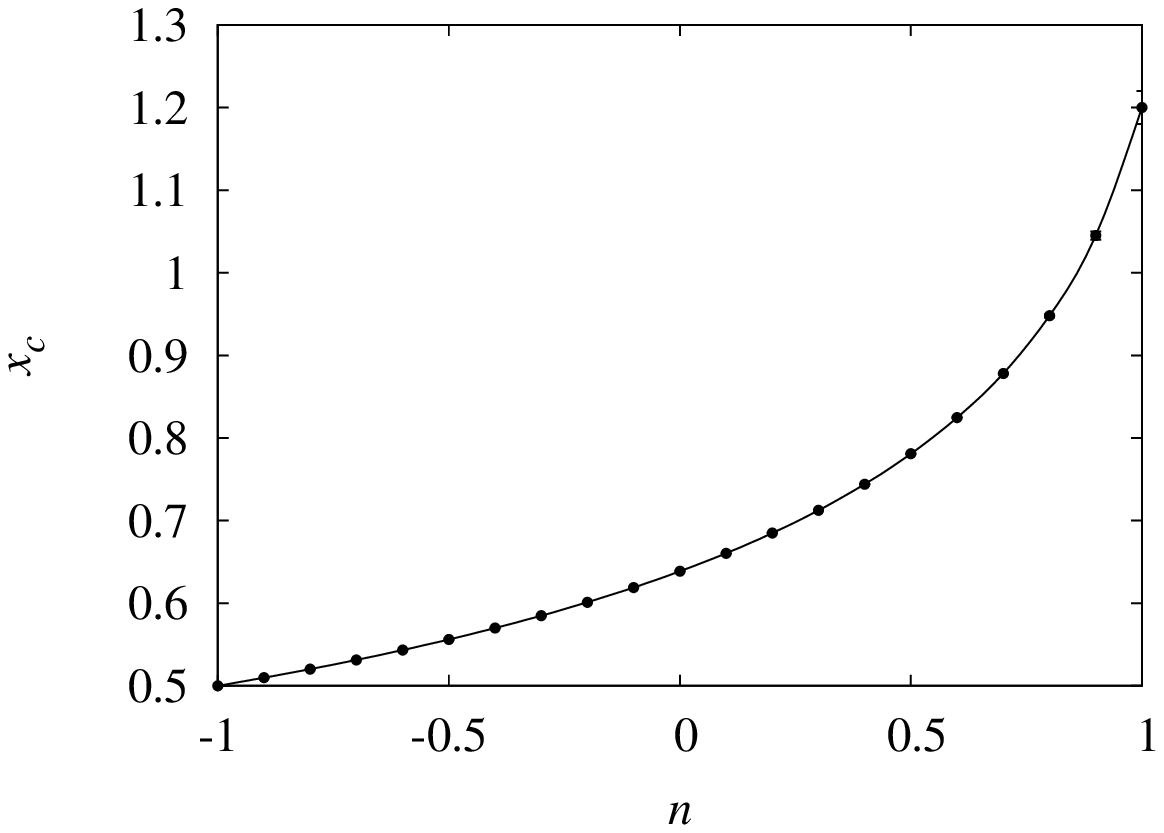}
\centering
\caption{
Location of the O($n$) transition point $x_c$ of the $y=z=0$ model, as a
function of the loop weight $n$. The error bars become visible only at
the right-hand side.
The line serves only as a guide to the eye.
}
\label{xc}
\end{figure}

Critical points of the $y=z=0$ model are shown in Fig.~\ref{xc}
for several values of $n$ in the range $-1 \leq n \leq 1$. The point
$x_c=0.5$ at $n=-1$ is exactly known; it is equivalent with the branch-0
point of Ref.~\onlinecite{BN} because the weight $z$ is redundant at
$n=-1$. The two orientations of the $z$-type vertex close a number of
loops differing by precisely 1, so that summation yields 0.
For $n \downarrow -1$ the magnetic gap closes, implying
$X_h \downarrow 0$, while $x_c$ approaches the precise value  $0.5$.

For $n=1$, Eq.~(\ref{Xhllp}) did not yield solutions; the scaled gaps
suggest marginal behavior, corresponding to an infinite-order transition.
Thus we expect $X_h =1/8$, which is consistent with the finite-size
results near the expected value of $x_c$. Thus we solved for $x$ in
\begin{equation}
X_h(x,L)=1/8
\label{Xhl}
\end{equation}
to obtain the critical point, presumably an infinite-order transition,
for $n=1$. Additional estimates were obtained using the transfer
matrix of the dual Ising representation and the scaling equation
$X_h(x,L)=X_h(x,L+2)$, also for even system sizes. 

\subsection{Conformal anomaly}
\begin{figure}
\includegraphics[scale=1.10]{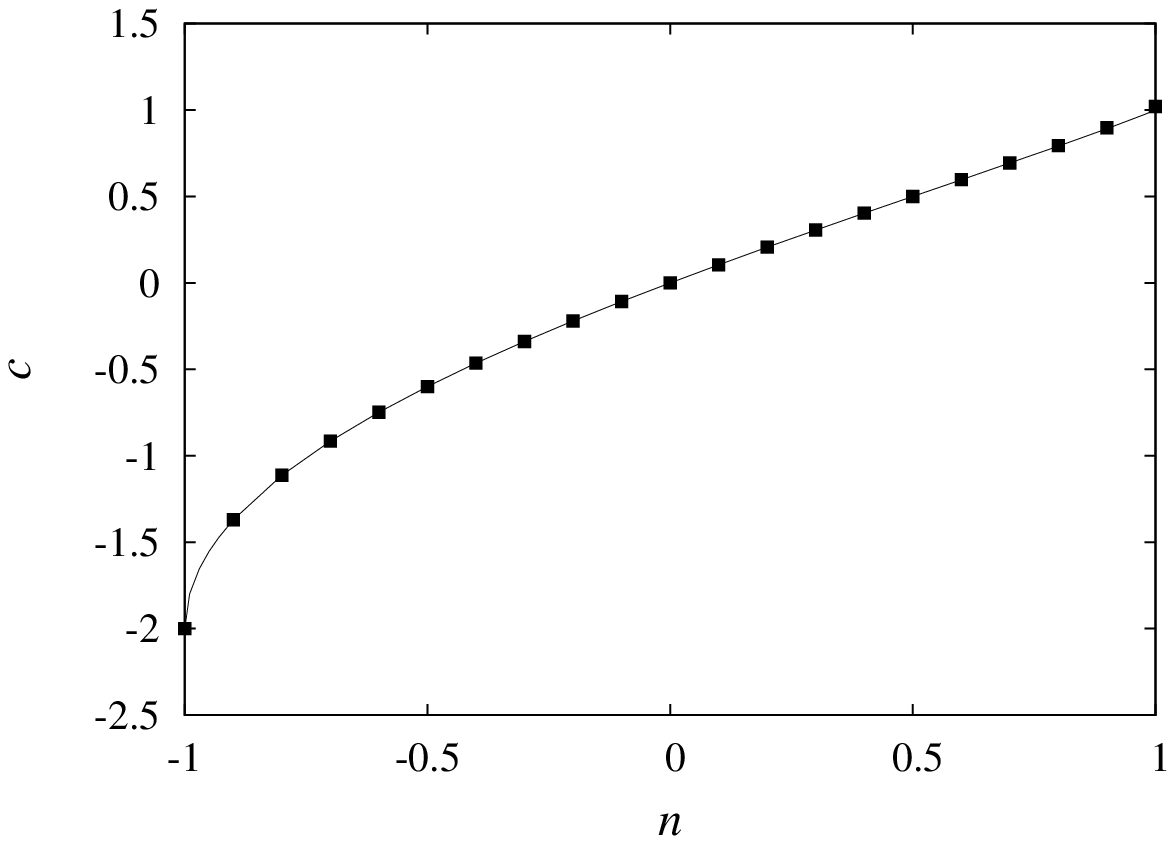}
\centering
\caption{
Conformal anomaly of the O($n$) model with $y=z=0$, versus the loop
weight $n$. These results, shown as data points,  do not agree with
the predictions for the generic O($n$) model.
Instead they agree well with the expected universal behavior of
the O(2$n$) transition, which is shown by the curve.
}
\label{ca}
\end{figure}

The conformal anomaly $c$ was numerically estimated as described in
Sec.~\ref{metan}. The results are shown in Fig.~\ref{ca}, together with
the Coulomb gas prediction Eq.~(\ref{caCG}) for the O(2$n$) model.
These data, which are, together with the $x_c$ estimates, also listed
in Table \ref{tabxcca}, show that the O(2n) universal classification is 
quite convincing, especially for $n<0.5$, thus confirming the picture
sketched in Ref.~\onlinecite{VJS}.
\begin{table*}[htbp]
\caption{Numerical results in the range $-1\le n\le 1$ for the critical
point $x_c(n)$  and the conformal anomaly $c$ of the O($n$) model with
$y=z=0$. Estimated numerical uncertainties in the last decimal place are
shown between parentheses.  We quote zero errors  in those cases where
all finite-size estimates coincide within numerical precision.
For comparison, we also include the exact conformal anomaly of the
generic O($2n$) model.}
\begin{center}
\begin{tabular}{|r||l||l|l|}
\hline
$n$  & $x_c(\rm num)$ &$c_{\rm num}$&$c_{\rm CG}$\\
\hline
-1.0 & 0.5        (0) & -2.0000  (1) & -2.0      \\
-0.9 & 0.5098053  (1) & -1.3700  (8) & -1.37061  \\
-0.8 & 0.5202394  (2) & -1.11330 (3) & -1.11331  \\
-0.7 & 0.5313745  (3) & -0.91570 (2) & -0.91572  \\
-0.6 & 0.5432954  (5) & -0.74835 (5) & -0.748403 \\
-0.5 & 0.556102   (1) & -0.59999 (2) & -0.6      \\
-0.4 & 0.569913   (1) & -0.46467 (3) & -0.464687 \\
-0.3 & 0.584873   (1) & -0.33898 (2) & -0.338996 \\
-0.2 & 0.601158   (1) & -0.22065 (2) & -0.220651 \\
-0.1 & 0.618984   (2) & -0.10805 (1) & -0.108051 \\
 0.0 & 0.638622   (2) &  0       (0) &  0        \\
 0.1 & 0.660420   (2) &  0.10443 (1) &  0.104434 \\
 0.2 & 0.684821   (2) &  0.20602 (2) &  0.206018 \\
 0.3 & 0.712433   (3) &  0.30543 (2) &  0.30541  \\
 0.4 & 0.74407    (1) &  0.40322 (5) &  0.403211 \\
 0.5 & 0.78090    (2) &  0.5002  (1) &  0.5      \\
 0.6 & 0.82465    (5) &  0.5968  (2) &  0.59639  \\
 0.7 & 0.8780     (2) &  0.694   (2) &  0.693093 \\
 0.8 & 0.948      (2) &  0.794   (3) &  0.791059 \\
 0.9 & 1.045      (5) &  0.897   (5) &  0.891858 \\
 1.0 & 1.20       (5) &  1.02    (1) &  1.0      \\
\hline
\end{tabular}
\end{center}
\label{tabxcca}
\end{table*}

\subsection{Critical exponents}
\subsubsection{Magnetic dimension}
\begin{figure}
\includegraphics[scale=1.10]{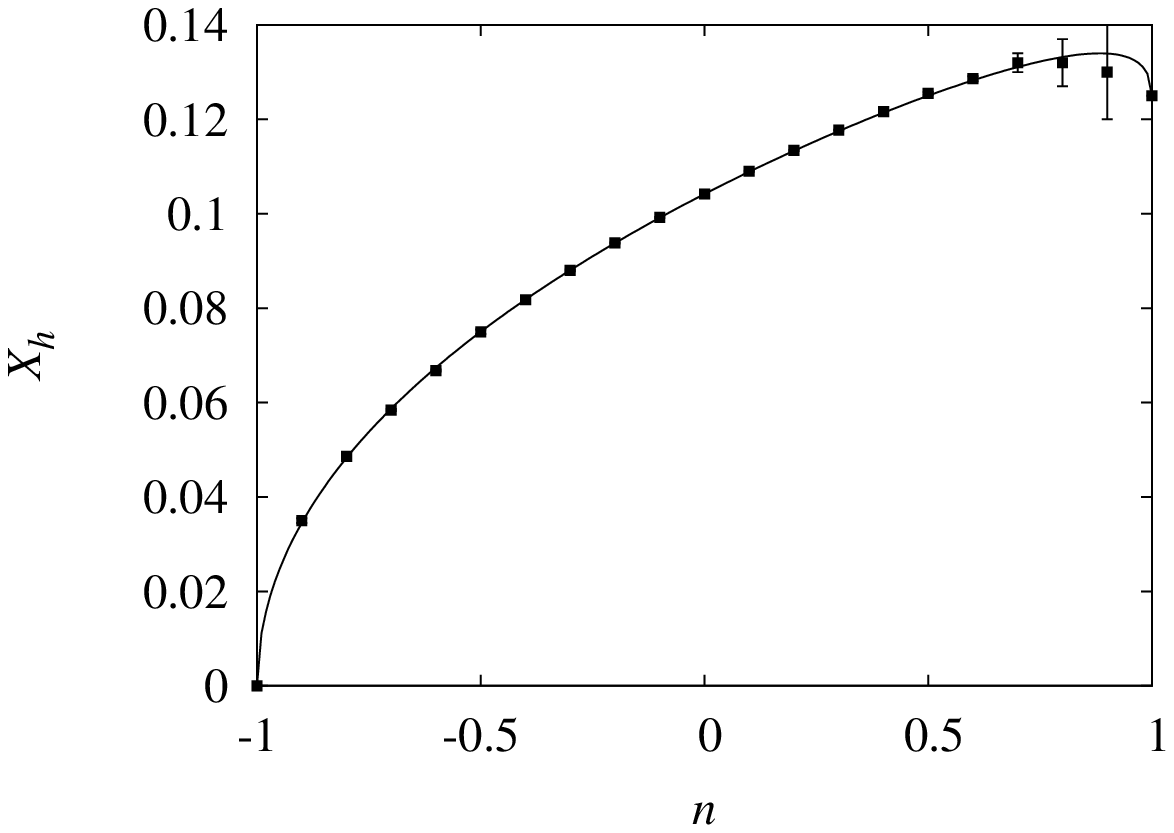}
\centering
\caption{
The data points display the results for the magnetic scaling dimension
of the O($n$) model with $y=z=0$, versus the loop weight $n$. The curve
shows the exactly known magnetic dimension of the generic O($2n$) model.
}
\label{xh}
\end{figure}

The numerical results for the magnetic scaling dimension
are shown as data points in Fig.~\ref{xh}. Since the eigenvalues
$\Lambda_0$ and $\Lambda_2$ coincide for $n=-1$, one has $X_h=0$
exactly.
The magnetic dimension of the generic O($2n$) critical point in two
dimensions, given by Eq.~(\ref{XthCG}), is included for comparison. 

\subsubsection{Temperature dimension}
\begin{figure}
\includegraphics[scale=1.10]{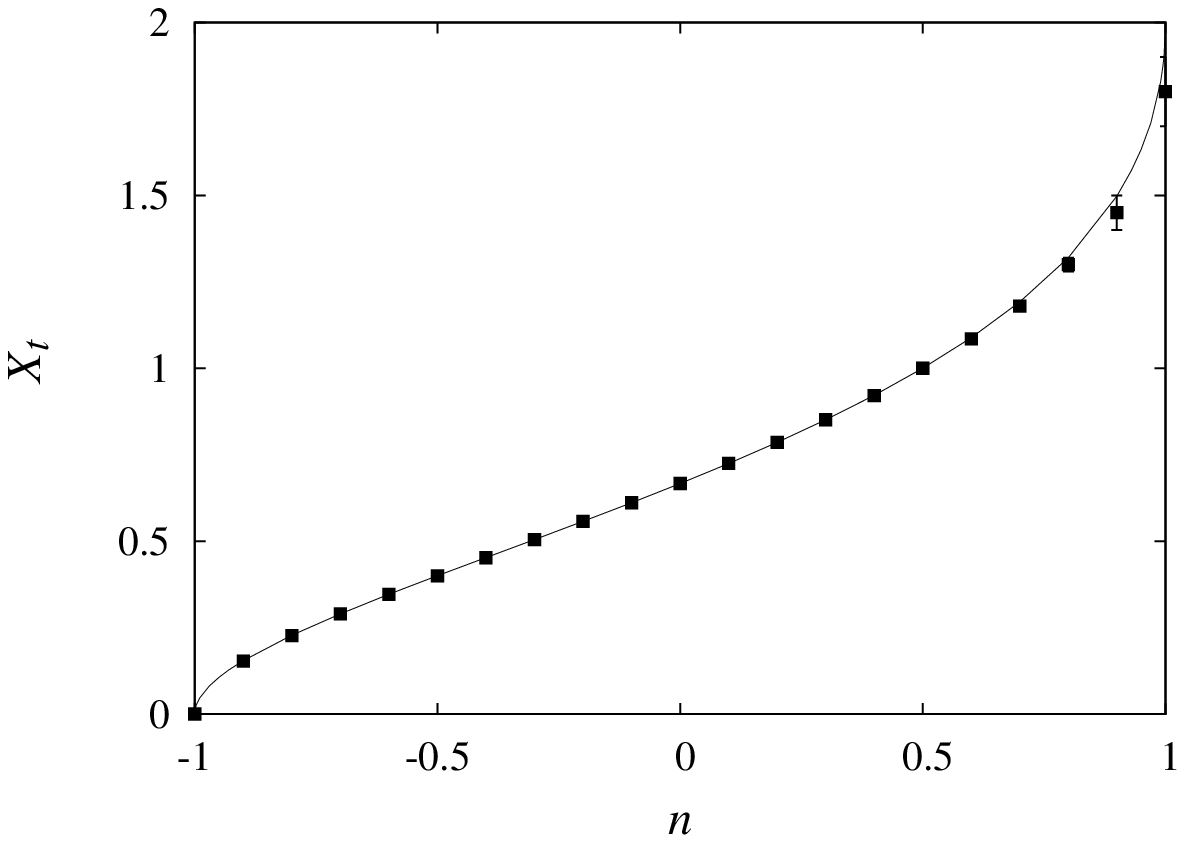}
\centering
\caption{
Temperature scaling dimension $X_t$ of the O($n$) model with $y=z=0$,
versus the loop weight $n$. The curve
shows the exactly known temperature dimension of the generic O($2n$) model.
}
\label{xt}
\end{figure}

The temperature dimension was obtained from the scaled thermal gaps
and the same methods of analysis as before. The results are shown
as data points in Fig.~\ref{xt}, together with the Coulomb gas
prediction Eq.~(\ref{XthCG}) for the O($2n$) model. 

\subsubsection{Topological dimension}
\label{topdim}
\begin{figure}
\includegraphics[scale=0.50]{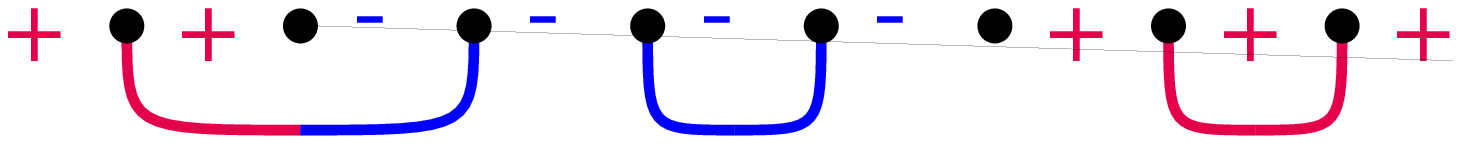}
\centering
\caption{(Color online)
An O($n$) loop connectivity with a two-colored loop. It corresponds
with a topological defect if $y=0$, because such a loop cannot be closed
if $y$-type vertices are absent. 
}
\label{topdef}
\end{figure}

\begin{figure}
\includegraphics[scale=1.10]{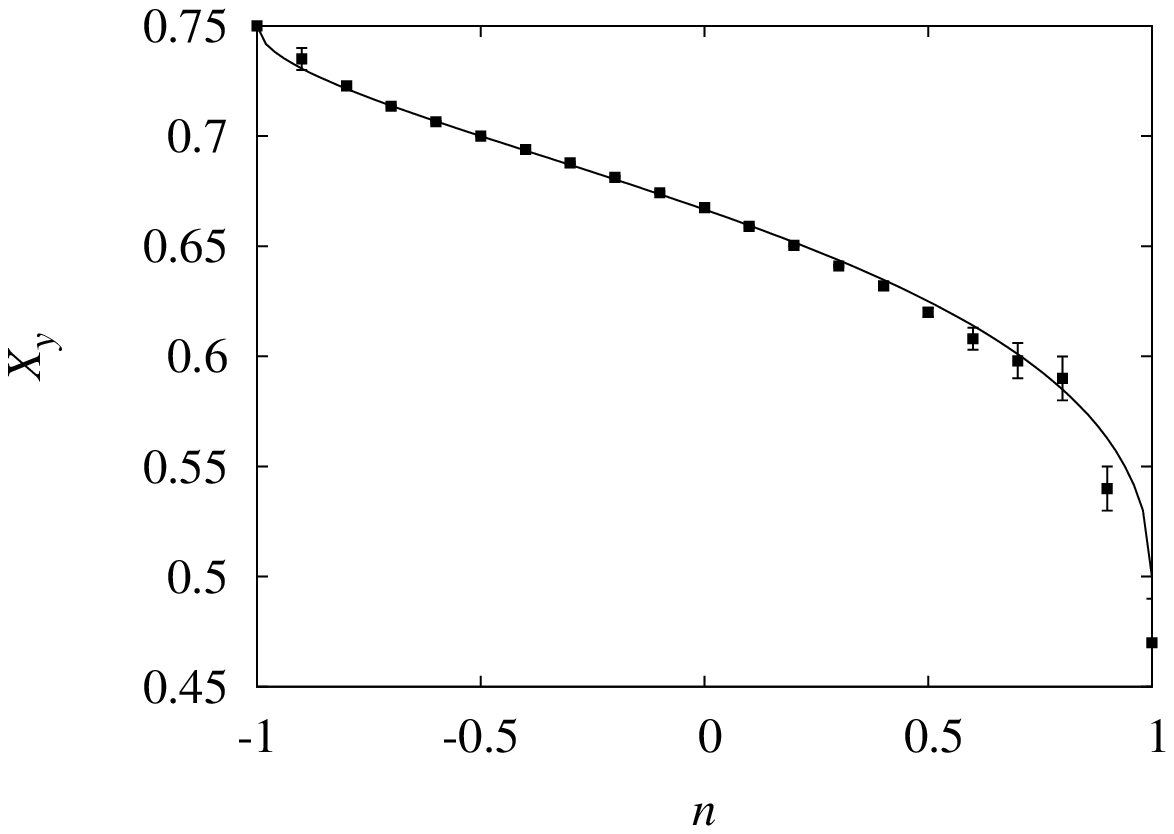}
\centering
\caption{
Topological dimension $X_y$ of the O($n$) model with $y=z=0$, 
versus the loop weight $n$.
The curve shows the Coulomb gas expression $1-1/2g$ as described in the text.
This dimension describes the probability that two remote points of the
O(2$n$) model lie on the same loop, and is equal to the conformal
dimension $2 \Delta_{0,1}$ of the latter model, and
to the magnetic exponent of the fully-packed O($2n$) 
model on the honeycomb lattice \cite{FPL}.
}
\label{xtopo}
\end{figure}

As argued in Ref.~\onlinecite{BN}, the Ising degree of freedom of a 
loop flips whenever a $y$-type vertex occurs. Closed loops must contain
an even number of these $y$-type vertices, which assume the role of
topological defects. In this work we exclude these vertices by choosing
$y=0$. But we can still study their effect by initializing a ``defective''
loop in which, {\em e.g.}, dangling bonds $k$ and $k+2$ are connected.
An example of such a connectivity, {\em i.e.}, the way in which the
dangling bonds are pairwise connected, is given in Fig.~\ref{topdef}.
This loop cannot be closed by transfer-matrix iterations if $y=0$.
The presence of such a defective loop defines another transfer-matrix sector,
whose leading eigenvalue we denote as $\Lambda_3$. Following the usual
procedure, we obtain the correlation length $\xi_y$ describing the
asymptotic behavior of the correlation function connecting two $y$-type
defects along the cylinder as
\begin{equation}
\xi_y^{-1}(L)=\ln (\Lambda_0/\Lambda_3) \, ,
\label{Xtop}
\end{equation}
from which the associated scaling dimension can be obtained by extrapolation
of the scaled gaps defined as
\begin{equation}
X_y(L)=\frac{L}{2 \pi \xi_y(L)} \, .
\end{equation}
The results for the scaling dimension of the $y$-type vertices
are shown in Fig.~\ref{xtopo}.
The results for $n<0.5$ are satisfactorily described by the simple formula
\begin{equation}
X_y=1- \frac{1}{2g} \, ,
\label{CGtop}
\end{equation}
where $g$ is the Coulomb gas coupling of the critical O($2n$) model,
{\em i.e.}, $\cos(\pi g)=-n$ with $1 \leq g \leq 2$.
For $n>0.5$ the differences become larger. We believe that this is due to
poor finite-size convergence associated with the proximity of a marginal
scaling field at $n=1$.
The numerical results for the scaling dimensions  of the O($n$) model 
with $y=z=0$ are summarized in Table \ref{tabxhty}, together with the
Coulomb gas values according to Eq.~(\ref{CGtop}) for the O(2$n$) model.
Again, the numerical result fits well  
in the O(2$n$) universality class, except in the neighborhood of $n=1$
where finite-size convergence is slow, and numerical uncertainties are
easily underestimated.

\begin{table*}[htbp]
\caption{Numerical results for the scaling dimensions $X_h$, $X_t$, and
$X_y$ of the O($n$) model with $y=z=0$. Estimated numerical uncertainties 
in the last decimal place are shown between parentheses. Zero errors
are shown where all finite-size estimates are zero within numerical
precision. For $X_{h}$,
$n=1$ we quote no error because the value $X_{h}=1/8$ was {\em assumed}
in the derivation of $x_c$. For comparison,
we also include the Coulomb gas results for the generic O($2n$) model.}
\begin{center}
\begin{tabular}{|r||l|l||l|l||l|l|}
\hline
$n$  & $X_{h,{\rm num}}$&$X_{h,{\rm CG}}$& $X_{t,{\rm num}}$&$X_{t,{\rm CG}}$
& $X_{y,{\rm num}}$&$X_{y,{\rm CG}}$\\
\hline
-1.0 & 0      (0) &      0    & 0      (0) & 0        & 0.7500 (1) & 0.75    \\
-0.9 & 0.0350 (5) & 0.0345037 & 0.153  (2) & 0.154669 & 0.735  (5) & 0.730666\\
-0.8 & 0.0486 (2) & 0.0482867 & 0.227  (1) & 0.228205 & 0.723  (1) & 0.721474\\
-0.7 & 0.0584 (1) & 0.0587088 & 0.2898 (1) & 0.28988  & 0.7135 (5) & 0.713765\\
-0.6 & 0.0668 (2) & 0.0674038 & 0.3460 (3) & 0.346271 & 0.7065 (5) & 0.706716\\
-0.5 & 0.0750 (5) & 0.075     & 0.3995 (5) & 0.4      & 0.7000 (1) & 0.7     \\
-0.4 & 0.0818 (5) & 0.0818165 & 0.4521 (5) & 0.452498 & 0.6939 (5) & 0.693438\\
-0.3 & 0.0880 (2) & 0.0880403 & 0.5045 (2) & 0.504717 & 0.6878 (5) & 0.68691 \\
-0.2 & 0.0938 (1) & 0.0937908 & 0.5573 (2) & 0.557391 & 0.6813 (5) & 0.680326\\
-0.1 & 0.0992 (1) & 0.099148  & 0.6112 (2) & 0.611163 & 0.6743 (5) & 0.673605\\
 0.0 & 0.1042 (1) & 0.104167  & 0.6668 (2) & 0.666667 & 0.6675 (5) & 0.666667\\
 0.1 & 0.1090 (1) & 0.108884  & 0.7252 (5) & 0.724581 & 0.6594 (5) & 0.659427\\
 0.2 & 0.1134 (1) & 0.113323  & 0.7859 (1) & 0.785698 & 0.6504 (5) & 0.651788\\
 0.3 & 0.1177 (2) & 0.117494  & 0.8509 (5) & 0.851006 & 0.641  (1) & 0.643624\\
 0.4 & 0.1216 (3) & 0.121394  & 0.921  (1) & 0.921819 & 0.632  (1) & 0.634773\\
 0.5 & 0.1255 (3) & 0.125     & 1.00   (1) & 1        & 0.620  (2) & 0.625   \\
 0.6 & 0.1286 (3) & 0.128262  & 1.085  (5) & 1.0884   & 0.608  (5) & 0.613949\\
 0.7 & 0.132  (2) & 0.131072  & 1.18   (1) & 1.19187  & 0.598  (8) & 0.601016\\
 0.8 & 0.132  (5) & 0.133192  & 1.30   (2) & 1.31996  & 0.59   (1) & 0.585005\\
 0.9 & 0.13   (2) & 0.133934  & 1.45   (5) & 1.49783  & 0.54   (1) & 0.562771\\
 1.0 & 0.125  (-) & 0.125     & 1.8    (1) & 2        & 0.47   (2) & 0.5     \\
\hline
\end{tabular}
\end{center}
\label{tabxhty}
\end{table*}
\section{Discussion}
\label{disc}
In the present loop model with $y=0$, the loops can in fact occupy
one of two sublattices. Together with the $n$ possible colors of
each loop, this leads in effect to a $2n$-fold degeneracy of the
loops. In this work, we provide an accurate confirmation of the
O($2n$) universal classification, in particular for $n<<1$. In the
neighborhood of $n=1$, the finite-size results are subject to poor
convergence, probably related to the irrelevant temperature exponent 
which is expected to become marginal for $n \uparrow 1$.

The phase diagram for $n=0.5$ shown in Sec.~\ref{phdiag} contains a part
that is difficult to resolve, in particular the part between $x=1/2$ and
$x=1$ of the critical line connecting to the multicritical point and 
forming the phase boundary of the Ising ordered phase. Near $x=0.7$, 
this transition is, because of its proximity to the O($n$)-type transition,
hard to distinguish from it, and it should be emphasized that the phase
diagram is not resolved here with certainty. If it is qualitatively
correct, then the line $z=x^2$ runs through two phase transitions,
implying the presence of an additional transition line for $y=0$ 
in Fig.~2 in Ref.~\onlinecite{FGB}.\\

While the present work is restricted to relatively small values of $n$, 
different phenomena are expected for large $n$ where the loops tend to
become small and behave as hard lattice-gas particles.
A line of transitions resembling the hard-square lattice-gas with
nearest-neighbor exclusion was located in Ref.~\onlinecite{FGB},
separating a dilute phase from one dominated by $z$-type vertices.
That result applies to the case $z=x^2$.  But also for $z=0$ and
sufficiently large $n$
one expects a transition when $x$ becomes larger, because one then
approximates the lattice gas with nearest- and next-nearest-neighbor
exclusion which displays a different type of transition \cite{Marq,FBN}.

The topological dimension $X_y$ defined in Sec.~\ref{topdim} does not
only describe the decay of the correlation function between two $y$-type
defects in the infinite plane as $r^{-2X_y}$, but it also determines the
crossover exponent $y_y=2-X_y$ describing the scaling  $y \to y'=b^{y_y}y$
under a rescaling by a scale factor $b$ near the $y=0$ fixed point. We
do indeed observe that the renormalization exponent $y_y$ is relevant
in the whole interval $-1 \leq n \leq 1$.

\acknowledgments
We thank Eric Vernier for informing us about the appearance of
Ref.~\onlinecite{VJS}. We are also indebted to Bernard Nienhuis for sharing
his valuable insights. H.B. is grateful for the hospitality extended 
to him by the BNU Faculty of Physics, where this work was performed.
This research was supported by NSFC Grants No.~11175018 and
No.~11447154, and by the Fundamental Research Funds for the Central
Universities (China).


\begin{thebibliography}{widest-label}
\bibitem{N}
B. Nienhuis, Phys. Rev. Lett. {\bf 49}, 1062 (1982);
J. Stat. Phys. {\bf 34}, 731 (1984).
\bibitem{Baxter}
R. J. Baxter, J. Phys. A {\bf 19}, 2821 (1986); J. Phys. A {\bf 20}, 5241
(1987).
\bibitem{BB}
M.~T. Batchelor and H.~W.~J. Bl\"{o}te, Phys. Rev. Lett.
{\bf 61}, 138 (1988); Phys. Rev. B {\bf 39}, 2391 (1989).
\bibitem{BNW}
M. T. Batchelor, B. Nienhuis and S. O. Warnaar,
Phys. Rev. Lett. {\bf 62}, 2425 (1989).
\bibitem{3WBN}
S. O. Warnaar, M. T. Batchelor and B. Nienhuis,
J. Phys. A {\bf 25}, 3077 (1992).
\bibitem{KNB}
Y. M. M. Knops, B. Nienhuis and H. W. J. Bl\"ote,
J. Phys. A {\bf 31}, 2941 (1998).
\bibitem{3WPSN}
S. O. Warnaar, P. A. Pearce, K. A. Seaton and B. Nienhuis,
J. Stat. Phys. {\bf 74}, 469 (1994).
\bibitem{VF}
V. A. Fateev, Sov. J. Nucl. Phys. {\bf 33}, 761 (1981).
\bibitem{S}
C. L. Schultz, Phys. Rev. Lett. {\bf 46}, 629 (1981).
\bibitem{PS}
J. H. H. Perk and C. L. Schultz, in
{\em Proc. RIMS Symposium on Non-Linear Integrable Systems},
edited by M. Jimbo and T. Miwa (World Scientific, 1983) p. 135;
and in {\em Yang-Baxter Equation in Integrable Systems},
edited by M. Jimbo (World Scientific, 1990) p. 326.
\bibitem{GNB}
W.-A. Guo, B. Nienhuis and H. W. J. Bl\"ote,
Phys. Rev. Lett. {\bf 96}, 045704 (2006).
\bibitem{BN}
H. W. J. Bl\"{o}te and B. Nienhuis, J. Phys. A {\bf 22}, 1415 (1989);
B. Nienhuis, Int. J. Mod. Phys. {\bf B4}, 929 (1990).
\bibitem{FGB}
Z. Fu, W.-A. Guo and H. W. J. Bl\"ote,
Phys. Rev. E {\bf 87}, 052118 (2013).
\bibitem{VJS}
E. Vernier, J. L. Jacobsen and H. Saleur,
{\em Dilute oriented loop models},
arXiv:1509.07768v2 [cond-mat.stat-mech] (2015).
\bibitem{BN82}
H.~W.~J. Bl\"{o}te and M.~P. Nightingale,
Physica A (Amsterdam) {\bf 112}, 405 (1982).
\bibitem{WGB} 
Y. Wang,  W.-A. Guo, and H. W. J. Bl\"ote,
Phys. Rev. E {\bf 91}, 032123 (2015).
\bibitem{BCN}
H. W. J. Bl\"{o}te, J. L. Cardy and M. P. Nightingale,
Phys. Rev. Lett. {\bf 56}, 742 (1986).
\bibitem{Affl}
I. Affleck, Phys. Rev. Lett. {\bf 56}, 746 (1986).
\bibitem{JCxi}
J. L. Cardy, J. Phys. A {\bf 17}, L385 (1984).
\bibitem{CG}
B. Nienhuis, in {\it Phase Transitions and Critical Phenomena}, Vol. 11,
eds. C. Domb and J. L. Lebowitz (Academic, London, 1987).
\bibitem{FSS}
For a review, see e.g. M. P. Nightingale in {\it Finite-Size Scaling and
Numerical Simulation of Statistical Systems}, ed. V. Privman (World
Scientific, Singapore 1990).
\bibitem{Lieb} 
E.~H. Lieb, Phys. Rev. Lett. {\bf 18}, 1046 (1967).
\bibitem{BWG} 
H. W. J. Bl\"ote, Y. Wang, and W.-A. Guo,
J. Phys. A {\bf 45}, 494016 (2012).
\bibitem{BBN}
H. W. J. Bl\"ote, M. T. Batchelor, and B. Nienhuis,
Physica A {\bf 251}, 95 (1998).
\bibitem{FPL}
H. W. J. Bl\"ote and B. Nienhuis,
Phys. Rev. Lett. {\bf 72}, 1372 (1994).
\bibitem{Marq}
H. C. Marques Fernandes, J. J. Arenzon, and Y. Levin, J. Chem. Phys. 
{\bf 126}, 11405 (2007).
\bibitem{FBN}
X. M. Feng, H. W. J. Bl\"ote and B. Nienhuis,
Phys. Rev. E {\bf 83}, 061153 (2011).
\end{thebibliography}
\end{document}